# Giulio Fermi's contributions to biophysics and molecular biology

Fabio Pichierri

*Department of Applied Chemistry, Graduate School of Engineering, Tohoku University,*

*Aoba-yama 6-6-07, Sendai 980-8579, Japan*

*E-mail address:* fabio@che.tohoku.ac.jp

[v.1, January 7, 2018]

## Abstract

This paper presents a comprehensive list of the scientific articles of Giulio Fermi (1936-1997), son of the Italian-American physicist Enrico Fermi, published between 1962 and 1997. The initial research activity of Giulio was concerned with virology and biological cybernetics while, from 1975 onward, his work was completely devoted to protein crystallography. The crystallographic research was carried out in collaboration with Nobel laureate Max Perutz at the Medical Research Council (MRC) Laboratory of Molecular Biology in Cambridge (United Kingdom). A short biography of Giulio (Judd) Fermi appears inside John Finch's book "A Nobel Fellow on Every Floor: A History of the Medical Research Council Laboratory of Molecular Biology" published by the MRC in 2008.

1. Introduction

Enrico Fermi (1901-1954) [1-6], the famous Italian-American physicist who displayed the rare ability of performing both experimental and theoretical research at equal levels of originality and creativity, received the Nobel Prize for Physics in 1938 *"for his demonstrations of the existence of new radioactive elements produced by neutron irradiation, and for his related discovery of nuclear reactions brought about by slow neutrons"* [7]. When the award ceremonies in Stockholm were over, after paying a short visit to Niels Bohr in Copenhagen, Enrico and his family fled to America on board of the RMS Franconia [3]. Once arrived in New York, Enrico told his wife Laura (1907-1977) that they had thus established the American branch of the Fermis [1].

Along with his wife there were two children, daughter Nella (1931-1995) and son Giulio (1936 -1997). Giulio, named in honor of Enrico's elder brother who passed away in 1915 while he was still a teenager, was initially attracted to mathematics (Master of Arts at Princeton University in 1957) but eventually studied biophysics and molecular biology at the University of California, Berkely, obtaining a Ph. D in 1961 under the guidance of Gunther S. Stent (1924-2008) [8]. After a postdoc in Europe at the Max Planck Institute (MPI) for Biology (now MPI for Biological Cybernetics) in Tubingen with Werner E. Reichardt (1924-1992) and a brief parenthesis as system analyst back in the United States, in 1971 Giulio joined the Medical Research Council Laboratory of Molecular Biology (MRC-LMB) in Cambridge (UK) under the wing of Nobel laureate Max Perutz (1914-2002) [9]. Here he performed research work on the structure and function of proteins, particularly on hemoglobin (Hb), the oxygen transporter found on the red blood cells of vertebrates [8]. Giulio (co)authored several scientific papers (vide infra) in the Journal of Molecular Biology (JMB) as well as in other biological or biochemical journals.

After more than twenty years since Giulio's death, I decided to collect his scientific papers in a list which provides an overview of his scientific ouvre. The majority of scientific publications in the list below (section 3) are available *on-line* in the corresponding journals' home-pages. To the best of my knowledge, apart for a short biography that appears inside John Finch's book [8], the scientific contributions of Giulio (Judd) Fermi have not been as yet summarized.

2. Methods

The Web of Science database [10] was accessed several times throughout 2017. To avoid confusion with the papers listed in the reference list placed at the end of this manuscript, Giulio Fermi's papers are numbered with Roman numerals. Along with the journal papers the list also comprises a paper contributed for a book chapter (XXXI) and an Atlas coauthored with Max Perutz (VII). Hereafter follows the list of scientific publications of Giulio Fermi while section 4 provides some comments about the research results of selected papers.

3. Bibliography of Giulio Fermi

I. G. Fermi, G.S. Stent, Protein Synthesis and the Onset of Intracellular Bacteriophage Growth ID. Replication of the Genome of a Secondary Phage,

University Press, Oxford (1993).

4. Comments

As it appears from Table 1, the majority of Giulio's papers (12) were published in the Journal of Molecular Biology (JMB). His first paper (I), coauthored with Gunther Stent, appeared in JMB in 1962. In this study the authors infected Escherichia coli (E.coli) bacteria with a primary T4 bacteriophage and allowed the synthesis of proteins to take place for a short time interval of 7 minutes. Next, they infected E.coli with a secondary T4 bacteriophage different from the first one and incubated the material for 50 minutes. The authors then observed that the protein formed by the primary T4 phage was used in the replication of the genetic material of the secondary T4 phage thereby concluding that such proteins are not specific to the phage employed. The follow-up study (II), published in the same year in the German journal Zeitschrift fiir Vererbungslehre, was concerned with the effect of E.coli infection by multiple viruses and its difference with respect to the single particle infection.

Table 1. Scientific journals containing Giulio Fermi's articles.

| Journal name | Nr. |
| --- | --- |
| J. Mol. Biol. | 12 |
| J. Biol. Chem. | 4 |
| P.N.A.S. | 4 |
| Biochemistry | 3 |
| Others | 10 |

The third paper (III) summarizes Giulio's work done at the Max Planck Institute (MPI) for Biology (now MPI for Biological Cybernetics) in Tubingen with Werner E. Reichardt. The paper, published in the journal Kybernetik (later renamed Biological Cybernetics) and written in German language (with the sole exception of the abstract which is written in English), is concerned with the optomotor reaction of the housefly *Musca domestica*. A single fly was fixed on a pin and the torque resulting from the movement of its wings was recorded by a rotating magnetic motor. Thus the authors were able to infer on the brightness of the light signals captured by the fly's eyes photoreceptors.

The fourth paper (IV), published in JMB in 1975, marks the beginning of Giulio's research in the field of biological crystallography (biocrystallography). This paper, where Giulio appears as the sole author, is devoted to the atomic structure of Human deoxyhemoglobin, the protein that transports molecular oxygen within the body, from the lungs to the tissues. Besides Human hemoglobin and its variants, such as hemoglobins Warsaw (paper XXIII) and Aalborg (paper XXVIII), crystallographic studies were also concerned with hemoglobins extracted from birds (paper XXV) and Antarctic fishes (papers XXVII and XXXIV). Interestingly, one paper (XXXII), published in the journal Structure, was concerned with the atomic structure of dimeric antithrombin. The crystallographic work on hemoglobins is summarized in a chapter (paper XXXI) written for a book edited by Koetzle et al. in 1993. Finally, in 1981 Giulio Fermi coauthored with Perutz an Atlas of Molecular Structures in Biology focused on their structural studies on hemoglobin and myoglobin (VII).

As far as the coauthors of Giulio's papers are concerned, paper VI contains the name of Robin Perutz, son of Max Perutz and Gisela Peiser (1915-2005). Robin is now a professor of inorganic chemistry at the University of York (UK). Other notable coauthors besides Max Perutz are: Maurizio Brunori (paper XXXV), now emeritus professor of Chemistry and Biochemistry at the University of Rome La Sapienza, former president (1990-1993) of the International Union of Pure and Applied Biophysics (IUPAB) [11], and from 2017 the vice president of the Accademia Nazionale dei Lincei (Rome) [12]; Kiyoshi Nagai (papers XVII, XXIX, and XXXIII), of the MRC-LMB in Cambridge whose group is currently engaged in disentangling the structure and dynamics of the spliceosome [13]; and Jeremy Tame (papers XXV, XXVI, and XXXIII) currently a professor of structural biology at Yokohama City University in Japan.

  A search of the Protein Data Bank (PDB) [14] produced nine entries for a query with the name "Fermi, G.": 1GLI, 1HBH, 1ANT, 1HDA, 1NIH, 1PBX, 2HHB, 3HHB, and 4HHB. The structure of 2HHB is depicted in Figure 1.

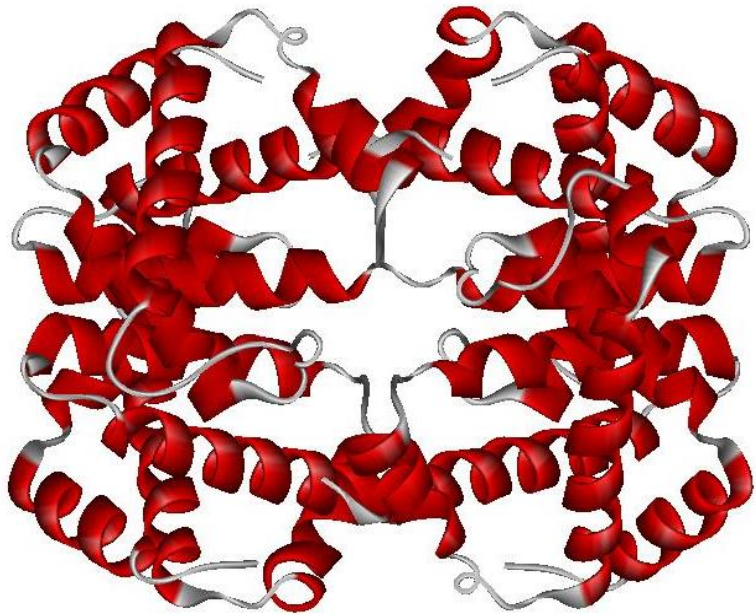

**Figure 1**. Solid ribbon depiction of the crystal structure of human deoxyhemoglobin (hetero-tetramer); PDB id: 2HHB; Ref.: G. Fermi, M. Perutz, B. Shaanan, R. Fourme, J. Mol. Biol. 175 (1984) 159-174.

## 5. Conclusions

This paper collects Giulio Fermi's scientific publications thus providing, for the first time, an overview of his research activities in biophysics and biological crystallography. From 1975 onward, most of his work was concerned with the determination of the molecular structure of hemoglobins using the technique of x-ray diffraction crystallography.

## Acknowledgements

I thank the Tohoku University Library for access to the Web of Science database as well as to specific journals mentioned in this work.